\documentclass[aps,prl,twocolumn,groupedaddress]{revtex4}

\usepackage{graphicx}
\usepackage{relsize}

\begin{document}               

\def\be{\begin{equation}}
\def\ee{\end{equation}}
\def\bd{\begin{displaymath}}
\def\ed{\end{displaymath}}
\def\ba{\begin{array}}
\def\ea{\end{array}}
\def\babar{\mbox{\slshape B\kern-0.1em{\smaller A}\kern-0.1em
    B\kern-0.1em{\smaller A\kern-0.2em R}}$\; $}
\def\232{D$_{sJ}^{*}(2317)^+$}

\title{Implications of a DK Molecule at 2.32 GeV} 
\author{T.Barnes} 
\email{tbarnes@utk.edu} 
\affiliation{Physics Division, 
Oak Ridge National Laboratory, Oak Ridge, TN 37831-6373, USA\\ 
Department of Physics and Astronomy, University of Tennessee,\\ 
Knoxville, TN 37996-1200, USA} 
\author{F.E.Close} 
\email{F.Close1@physics.ox.ac.uk} 
\affiliation{ Department of Theoretical Physics, University of Oxford,\\ 
Keble Rd., Oxford OX1 3NP, UK} 
\author{H.J.Lipkin} 
\email{lipkin@hep.anl.gov} 
\affiliation{Department of Particle Physics, Weizmann Institute of Science, 
Rehovot, Israel\\ 
School of Physics and Astronomy, Raymond and Beverly Sackler 
Faculty of Exact Sciences\\ 
Tel Aviv University, Tel Aviv, Israel\\ 
High Energy Physics Division, Argonne National Laboratory,\\ 
Argonne, IL 60439-4815, USA}


\begin{abstract}
We discuss the implications of a possible quasinuclear 
DK bound state at 2.32 GeV. Evidence 
for such a state was recently reported in D$_s^+\pi^o$
by the \babar Collaboration. 
We first note that a conventional quark model $c\bar s$ assignment 
is implausible, and then consider other options involving multiquark 
systems. An I=0 $c\bar s n \bar n$ baryonium assignment is one possibility. 
We instead favor a DK meson molecule assignment, which can account for the 
mass and quantum numbers of this state. 
The higher-mass scalar $c\bar s$ state expected at 2.48~GeV 
is predicted to have a very large DK coupling, which would encourage 
formation of an I=0 DK molecule. Isospin mixing is expected in 
hadron molecules, and a dominantly I=0 DK state with some 
I=1 admixture could explain both the narrow total width 
of the 2.32 GeV state as well as the observed decay to D$_s^+\pi^o$. 
Additional measurements that can be used to test this and related 
scenarios are discussed. 
\end{abstract}

\maketitle

\section{Introduction}

The \babar Collaboration recently reported 
a narrow state near 2.32 GeV, 
known as the D$_{sJ}^{*}(2317)^+$, decaying to 
D$_s^+\pi^o$ \cite{Bab03}. 
The observed width is consistent with experimental resolution, 
which gives a limit of $ \lesssim 10$~MeV for the total width.
For reference purposes 
we show the new state at 2.32~GeV in Fig.1,
together with the Godfrey-Isgur-Kokoski predictions 
for the spectrum of $c\bar s$ mesons \cite{God85,God91},
DK thresholds, 
and the experimental spectrum of charm-strange states 
\cite{PDG02}. 

One might {\it a priori} consider a new 
resonance observed in D$_s^+\pi^o$
in this mass region
to be a candidate $c\bar s$ quark model state,  
decaying to D$_s^+\pi^o$ through 
an isospin-violating strong decay. 
Since the D$_{sJ}(2573)$ is already well established as a plausible 
$^3$P$_2$ $c\bar s$ candidate, the only available assignment 
would be the $^3$P$_0$ D$^*_{s0}$ level. 

\begin{figure}
\includegraphics[width=0.4\textwidth, bb=  61 84 717 525, clip=]{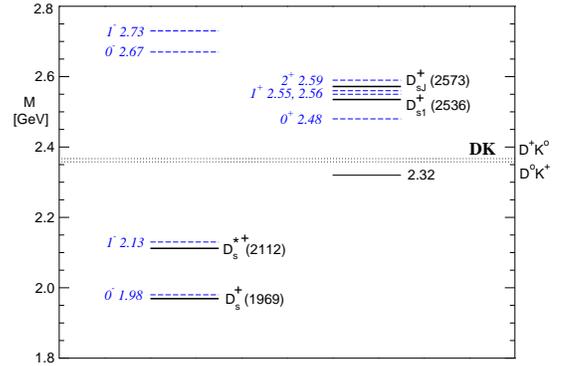} 
\caption{The experimental (solid) and theoretical (dashed) spectrum of 
$c\bar s$ mesons. DK thresholds and the 2.32~GeV BaBar state are also shown.} 
\label{fig:bc_fig1} \end{figure}

Identification of the 2.32~GeV signal with the $^3$P$_0$
$c\bar s$ quark model state 
appears implausible for two reasons.
First, the mass predicted by Godfrey and Isgur for this
$c\bar s$ state is 2.48~GeV,
160~MeV higher than the BaBar state.  
Second, as the scalar $^3$P$_0$ $c\bar s$ belongs to the $j=1/2$ 
heavy quark symmetry doublet, both the $^3$P$_0$ $c\bar s$
and its D$_{s1}$ partner are
expected to be much broader than the states in the $j=3/2$ doublet. 
The $j=3/2$ doublet is usually
identified with the rather narrow D$_{sJ}(2573)$ and 
D$_{s1}(2536)$, which have experimental total widths of 
$15{+5\atop -4}$~MeV and
$ < 2.3$~MeV ($90\%\;  c.l.$) respectively.
In contrast, a total width of 270-990~MeV (depending on the decay 
model assumed) was predicted for the $^3$P$_0$ $c\bar s$ scalar 
by Godfrey and Kokoski \cite{God91}, 
assuming a mass of 2.48~GeV.

\section{Multiquarks options}

Assuming that the new 2.32~GeV state is being observed in a strong 
or electromagnetic decay 
to D$_s^+\pi^o$, it must at least possess $c$ and $\bar s$ quarks. 
Given the implausibility of identifying this signal with a $c\bar s$ 
quark model state, as discussed above, we are led to the 
consideration of states with additional valence quarks. 
The proximity to the lightest $c\bar s$ states suggests the first available 
color-singlet combination, $cn\bar s \bar n$ (where $n$ generically 
represents either of $u,d$).

Four-quark states \cite{Jaf77} may be classified as 
``baryonia" if the spatial wavefunction
is well described as a single multiquark cluster, 
or ``molecules" if they
are dominantly quasinuclear, weakly bound pairs of $q\bar q$ mesons. 
A subcategory of baryonia are the 
``heavy-light" systems, which possess a heavy pair and
a light pair, such as $QQ\bar n \bar n$ or $Q n \bar Q \bar n$. These 
states are interesting because the heavy pair 
is spatially localized and should be 
dominantly in a particular color state \cite{Ric99}.
The DK system was previously suggested as a possibility for 
four-quark bound states of both baryonium and molecular 
types by Lipkin \cite{Lip77,Lip86}
and Isgur and Lipkin \cite{Isg81}. 
  
For our initial discussion 
we will treat these as distinct categories of multiquark states,
although this is clearly a rather qualitative distinction. 
One may actually find significant amplitudes 
for both types of spatial configurations in some resonances;
see for example the discussion of the $f_0/a_0(980)$
in Ref.\cite{Clo02}.

\subsection{Baryonia}

Baryonia composed of light quarks do not require an interaction to 
dissociate into light meson pairs; this is known as ``fall-apart" decay. 
This effect implies that light baryonia may not exist as 
resonances at all, or if they do exist they are expected to be 
extremely broad \cite{Jaf77}. 
For this reason
it would be difficult to identify the
2.32~GeV BaBar signal with an I=1 $cn\bar s \bar n$ baryonium 
state; it would have a fall-apart decay to D$_s^+\pi$, and so should
be extremely broad or nonresonant.

An I=0 $cn\bar s \bar n$ baryonium state is a more interesting
possibility; there is no accessible fall-apart mode, since DK 
does not open until 2.36~GeV. 
The channel D$_s^+\pi^o$ would be open to isospin-violating transitions, 
but this coupling might be sufficiently weak to allow an
I=0 $cn\bar s \bar n$ cluster to appear as a resonance.    
If we assume that the 2.32~GeV signal is indeed an I=0 $cn\bar s \bar n$ 
baryonium, other I=0 $cn\bar s \bar n$ states with different 
angular quantum numbers may also lie below DK threshold. 
If baryonium models instead predict no other $cn\bar s \bar n$  
states below 2.36~GeV,  
it may prove difficult to distinguish between I=0 $cn\bar s \bar n$ 
baryonium and DK molecule assignments. The proximity of 
the DK threshold to
2.32~GeV is of course an argument in favor of a 
DK molecule, since this would be accidental for a baryonium state.

If attractive interquark forces do form an 
I=0 $cn\bar s \bar n$ baryonium bound state at 2.32~GeV, 
one might also anticipate I=1 and $cs\bar n \bar n$ partners
nearby in mass. A natural spin-parity I=1 $cn\bar s \bar n$  
baryonium above
2.25~GeV would have a fall-apart mode to D$_s^+\pi$ and hence 
should be very broad or nonresonant. 
The presence of such a hypothetical resonance 
might be observable in 
$e^+e^-$ annihilation (see our subsequent discussion).
In contrast, in the DK molecule scenario an I=1 bound state is less likely,
as we shall explain in the following section.

Exotic-flavor $cs\bar n \bar n$ 
baryonium partner states would provide dramatic support for the 
baryonium picture. If these states were
below 2.36~GeV 
(D$\bar {\rm K}$ threshold) they 
would only decay weakly (see subsequent discussion of baryonia). 
If the baryonium scenario is correct, $cs\bar n \bar n$ states
should be produced in $e^+e^-$ at a rate comparable to the
BaBar state.

\subsection{Molecules}

Hadronic molecules are systems that to a good approximation 
are weakly bound states of color-singlet hadrons. Nuclei and 
hypernuclei are the most familiar examples of these states, 
although there are several 
often-cited candidates for meson-meson molecules, 
notably the $f_0(980)$ and $a_0(980)$ \cite{Clo02,Wei90} and $\psi(4040)$ 
\cite{Nov76,Nov78,Geo77,Iwa80}, and at least one meson-baryon candidate, 
the $\Lambda(1405)$ \cite{Dal60,Sak60}. 

The best studied candidates for meson-meson molecules are the 
$f_0(980)$ and $a_0(980)$, which are widely believed to have 
large or perhaps dominant K$\bar{\rm K}$ components. This sector 
of the quark model was studied in detail 
by Weinstein and Isgur \cite{Wei90}, 
who concluded that conventional quark model 
forces gave rise to attractions in the I=0 and I=1 K$\bar{\rm K}$
channels that are sufficiently strong to form bound states.   
Their conclusions regarding the nature of these attractive forces 
may also be relevant for the 2.32~GeV BaBar signal, 
as the K$\bar{\rm K}$ and DK systems
share several important features.

Weinstein and Isgur found that the dominant attraction in the S-wave 
K$\bar{\rm K}$ system arose from level repulsion between 
the low-mass K$\bar{\rm K}$ continuum and scalar $q\bar q$ states. 
The $q\bar q$ scalars were assumed to lie near 1.3~GeV, and to have 
strong couplings to two-pseudoscalar channels. These scalar mesons 
play a crucial role as ``shepherd states" which drive the two-meson 
continuum into bound states just below threshold. Additional non-resonant 
forces between pseudoscalar meson pairs were found by 
Weinstein and Isgur in their variational study of the $s n \bar s \bar n$ 
system \cite{Wei90}; these were subsequently identified as arising 
mainly from the one-gluon-exchange contact spin-spin interaction, 
which dominates constituent-interchange scattering \cite{Bar92}. 
In the final Weinstein-Isgur paper this interaction couples several 
two-pseudoscalar channels, and provides additional attraction in both 
K$\bar{\rm K}$ channels. 

Since the residual forces that bind hadrons 
into molecules are relatively weak and short-ranged, 
simple qualitative signatures for hadron-pair molecules 
can be abstracted from the Weinstein-Isgur results.    
These are 

\vskip 0.5cm
\noindent
1) J$^{\rm PC}$ and flavor quantum numbers of an L=0 hadron pair, 

\vskip 0.2cm
\noindent
2) a binding energy of at most about 50-100 MeV, 

\vskip 0.2cm
\noindent
3) strong couplings to constituent channels, and

\vskip 0.2cm
\noindent
4) anomalous electromagnetic couplings relative to expectations for 
a quark model state.

\vskip 0.5cm
\noindent
The justification for each of these proposed molecule signatures  
is discussed in Ref.\cite{Bar94}, together with a review
of earlier experimental candidates.

\section{A DK Molecule?}

\subsection{DK and molecule signatures}

The 2.32~GeV 
BaBar signal appears to be an obvious candidate for a scalar DK molecule,
since what is known about this state 
satisfies the first two of the molecule signatures quoted above. 
First, 
the (assumed strong or electromagnetic) decay to D$_s^+\pi^o$ 
implies natural spin-parity, 
so J$^{\rm P} = 0^+$ is allowed. 
(Note further that for strong decays 
the combined observation in D$_s^+\pi$ and absence in
D$_s^{*+}\pi$ would uniquely select J$^{\rm P} = 0^+$.)
Second, the DK thresholds are 
m(D$^o$K$^+$) = 2358~MeV,
m(D$^o$K$^o$) = 2362~MeV,
m(D$^+$K$^+$) = 2363~MeV and
m(D$^+$K$^o$) = 2367~MeV, so a DK molecule at 2.32~GeV would have a plausible 
binding energy of $\approx 40$~MeV. 
The third signature is more problematic since the only open 
strong mode for a J$^{\rm P} = 0^+$ DK molecule is D$_s^+\pi$, 
and this may be an isospin-suppressed decay; 
this will be discussed subsequently. 
The final signature can be used as a test of the molecule assignment, 
through a 
measurement of D$_{sJ}^{*}(2317)^+  \to D_s^{*+}\gamma$; 
this E1 transition rate can be calculated 
for a $^3$P$_0$ $c\bar s$ quark model state at 2.32 GeV, 
which predicts 
$\Gamma_{\gamma {\rm D}_s^{*+}} \approx 2$~keV 
\cite{God03}.
If this is indeed a non-$c\bar s$ state, 
one would expect a rather different
rate for the E1 transition. 
This comparison is well known for $\phi \to \gamma f_0 / a_0 (980)$; 
the rate for a molecule was computed in Ref.\cite{Clo93}. 
The analogous computation for
a DK molecule would require knowledge of its coupling strength 
to both DK and
D$_s^{*+}$.

If this state is a DK molecule or a baryonium resonance, 
power counting rules \cite{Bro73}
imply that its elastic form factor should fall as 
$1/Q^6$, 
in contrast to the $1/Q^2$ expected 
for a ``normal" $c\bar{s}$ state.
At CLEO-c one could pair produce the open-charm meson 
states, including the BaBar state as well as conventional 
charmed quark meson pairs, near threshold. 
The anomalous $Q^2$-dependence of the 
exclusive channel cross section 
could then confirm its four-quark nature, 
or conversely, if established as a multiquark system, 
could provide a novel further test of the quark counting rules.
Note that at large $Q^2$ one would expect to see 
a weakened $1/Q^2$ dependence from 
the $c\bar{s}$ component of the BaBar state, 
which is expected to be present 
at some level due to mixing effects.

\subsection{Previous studies of the DK system}

Motivated by Jaffe's study of light baryonium states in the bag model 
and the suggested classification of light scalars as four-quark states 
\cite{Jaf77}, Lipkin \cite{Lip77} suggested that four-quark 
baryonium systems of the 
type $c\bar s n \bar n$ and $c s \bar n \bar n$ 
might also be observed as resonances. In the cluster wavefunctions 
tacitly assumed in this paper the dominant binding force was taken to be 
the one-gluon-exchange color magnetic force, as in the MIT bag model.  
Decay systematics of the various possible states were discussed, 
and it was noted that for masses between D$_s^+\pi $ 
and DK the I=1 $c\bar s n \bar n$ 
state ``$\tilde {\rm F}_{\rm I}$" could decay strongly to D$_s^+\pi $, 
but a pure I=0
$c\bar s n \bar n$ ``${\rm F}^+_x$"
would only have electromagnetic modes, such as
D$_s^+\pi^o $, D$_s^+\gamma\gamma $ and D$_s^+\pi^o \gamma$. 
Although the states were assumed to be baryonia,
the decay systematics apply to molecular bound states with the
same quantum numbers as well.

Isgur and Lipkin \cite{Isg81}
stressed the important distinction between four-quark baryonium clusters 
and hadronic molecules, and observed that 
the determination of which type of configuration best describes 
the ground state of a given bound system is a problem with
``no simple model-independent answer". The 980~MeV states are cited as 
examples near the molecular limit, ``just barely bound states of the 
K$\bar {\rm K}$ system". It is suggested that ``similar bound states of 
D$\bar {\rm K}$ and DK ..." (hence molecules rather than clusters) 
``...should exist near and possibly below the DK threshold". 
Assuming as in \cite{Lip77} that the dominant interaction 
is the color magnetic spin-spin hyperfine interaction, 
Isgur and Lipkin gave estimates of the masses of 
$c\bar s n \bar n$ and $cs \bar n \bar n$ systems relative to DK. 
Although their estimates 
find masses above DK threshold by 
205 and 140~MeV respectively, 
they argued that the smaller kinetic energies of charmed 
systems suggest that weakly bound DK and perhaps D$\bar {\rm K}$ 
molecules exist. 
The mode D$\bar {\rm K} \to 
\bar {\rm K}^o \bar {\rm K}^o$ was proposed for searches for 
a D$\bar {\rm K}$ molecule,
for example in ${\rm B} \to $ (D$\bar {\rm K}$)
${\rm K}^o \to (\bar {\rm K}^o \bar {\rm K}^o ) {\rm K}^o$.

In discussing early results for light multiquark 
systems one should note that 
Weinstein and Isgur \cite{Wei90} subsequently
found that
level repulsion against higher-mass $q\bar q$ states gave a larger 
attraction than the color magnetic interaction. This 
additional force 
will contribute to binding in the I=0 DK case, but not in I=1 DK or 
any D$\bar {\rm K}$ channel.

An additional development has been the realization that isospin 
mixing is important in molecular states, which was not appreciated 
in the early references. In particular this allows 
``isospin violating" strong decays from 
a dominantly I=0 DK molecule, as we shall discuss below.

Lipkin \cite{Lip86} has also considered four-quark systems containing both
heavy and light quark pairs, such as $cc\bar u \bar d$. For sufficiently
large heavy quark mass these systems take on a baryon-like spatial
configuration, with the two heavy quarks acting as a single heavy
antiquark. These heavy-light systems constitute a distinct category of
four-quark state, and for sufficiently large heavy quark mass are expected
to be strongly stable \cite{Lip86,Ric99}.
The Coulomb-like color electric attraction between the two heavy quarks
produces binding in this model, whereas the color-magnetic interaction is
inversely proportional to quark mass and so is 
neglected for the heavy quarks.
The strange quark is not heavy enough to produce a bound state 
in this heavy-light model; its color-magnetic interaction was crucial 
for binding in the other early studies \cite{Lip77,Isg81}

Ref.\cite{Lip86}
considered only heavy-light baryonia with identical heavy 
quarks, and concluded that $cc\bar u \bar d$ is probably not bound but
$bb\bar u \bar d$ may well be. Extending this approach to states with
nonidentical heavy quarks leads to the conclusion that $cs\bar u \bar d$
is not bound, but $bc\bar u \bar d$ may well be \cite{Lip03}. 
This state 
would
decay only weakly, either by b-quark decay into two charmed mesons or 
c-quark decay into
a B meson and a strange meson.  The corresponding signature 
in a vertex detector would be a secondary vertex with a 
multiparticle decay, one or two subsequent heavy quark decays, 
and either one or no tracks from the
primary vertex to the secondary.

\subsection{DK isospin and isospin mixing}

The isospin of the purported DK molecule is a nontrivial issue. 
Were isospin a good quantum number, the narrow width would suggest 
I=0; there are then no open strong modes, so the state would be very 
narrow, and the observed decay to D$_s^+\pi^o$ would be a suppressed 
isospin-violating 
transition. I=0 is also favored by the dominant 
molecule-binding mechanism found for K$\bar{\rm K}$ by 
Weinstein and Isgur, which is repulsion of the lower continuum 
against a higher-mass scalar $q\bar q$ state. For I=0 
we do have such a state, the $^3$P$_0$ $c\bar s$ D$_{s0}^*(2.48)$ 
of Godfrey and Isgur \cite{God85}, which was predicted 
by Godfrey and Kokoski \cite{God91} to have a 
very strong coupling to the DK continuum,
as required to induce binding.

In contrast, for a pure I=1 molecule there can be no DK attraction due to 
level repulsion against a $q\bar q$, since $c\bar s$ has I=0.
Binding might instead arise from
diagonal DK forces and 
repulsion against other two-meson channels, such as D$_s^{*+}\rho$. 
Note however that the diagonal DK interaction in I=1 should be weak, since 
constituent interchange is purely off-diagonal, 
$
(c\bar n)(n'\bar s)
\to
(c\bar s)(n'\bar n)
$.

The I=1 DK molecule option can be tested by searching for 
I$_z = \pm 1$ partner states. Assuming that the BaBar state is 
produced strongly, starting from $e^+e^-\to \gamma \to c\bar c$, 
the overall hadronic system would have I=0. Partitioning the 
final hadronic state as 
\be 
|{\cal F}\rangle_{\rm I=0} =
|{\rm DK}\rangle_{\rm I=1}\; \otimes \;|{\rm everything\ else}\rangle_{\rm I=1}
\ee
the CG coefficients in $0\subset 1\otimes 1$ imply that 
I, I$_z = 1,\pm 1$ partner DK 
states would each be produced at the same rate as an
I, I$_z = 1, 0$ DK molecule. The partner states would decay into 
D$_s^+\pi^{\pm}$ at the same isospin-allowed rate as the 
I, I$_z = 1, 0$ state. Thus one can test the possibility of an 
I=1 DK molecule quite easily by searching for D$_s^+\pi^{\pm}$ events 
at 2.32~GeV; if the BaBar state is I=1, one should see similar numbers 
of D$_s^+\pi^+$, D$_s^+\pi^-$ and D$_s^+\pi^o$ events. In contrast, 
if it is 
dominantly I=0, the signal in 
$e^+e^- \to ({\rm D}_s^+ \pi^o){\rm X}^-$
should greatly exceed 
that in
$e^+e^- \to ({\rm D}_s^+ \pi^+){\rm X}^{--}$
and
$e^+e^- \to ({\rm D}_s^+ \pi^-){\rm X}^o$;
naive
isospin rules predict that it should be
completely absent in the charged-pion reactions. 

Although the I=0 channel is favored theoretically
for DK molecule formation through
the Weinstein-Isgur mechanism, we emphasize that a
nominally I=0
DK molecule is actually expected to show significant 
isospin mixing with the
$|{\rm I, I}_z\rangle = |1, 0\rangle$ DK basis state.
Indeed, this isospin mixing 
is one of the characteristic features of molecules
\cite{Ach81, Bar85}, and has probably been observed 
in the $f_0/a_0$(980) states
(see for example \cite{Lin97} and \cite{Clo00}). 
The reason for this 
isospin mixing is that hadrons within an
isomultiplet typically have $\approx 5$~MeV mass splittings, 
which is significant on the scale of molecule binding energies. 

We can illustrate this effect using a 
simple two-state model. Consider a Hamiltonian 
that couples the nondegenerate two-meson states
$|{\rm D}^+ {\rm K}^o\rangle = |{\rm A}\rangle$
and
$|{\rm D}^o {\rm K}^+\rangle = |{\rm B}\rangle$
through an I=0 s-channel interaction,

\be
{\rm H} = 
\left[
\ba{cc} 
m_0 + \frac{1}{2}\delta m  &      \\ 
                           &  m_0 - \frac{1}{2}\delta m  
\ea
\right]
+\frac{v}{2}
\left[
\ba{cc}
-1 &  \phantom{-}1    \\
 \phantom{-}1 & -1      
\ea
\right]\ .
\ee 
In the weak coupling limit ($v << \delta m$) the ground state approaches  
$|\psi_0\rangle = |{\rm B}\rangle =
(|1,0\rangle - |0,0\rangle)/\sqrt{2}$, 
a linear combination
of I=0 and I=1 states with equal weight 
(thus maximally violating isospin). 
For very large coupling ($v >> \delta m$) isospin symmetry is restored, 
and the system approaches a pure
I=0 ground state, 
$|\psi_0\rangle = (|{\rm A}\rangle - |{\rm B}\rangle)/\sqrt{2} = 
|0,0\rangle $, with energy E$_0 = m_0 - v$. For moderately large coupling, 
as is presumably appropriate here, the ground state is close to 
I=0 but has a significant I=1 component, 
\be
|\psi_0\rangle = |0,0\rangle - \frac{\delta m}{2v} |1,0\rangle
+ O\Big(\frac{\delta m }{v}\Big)^2 \ .
\ee 

In DK there is a rather
large splitting between free two-meson states, 
\be
\delta m = m( {\rm D}^+ {\rm K}^o ) -
m( {\rm D}^o {\rm K}^+ ) = 9.3 \pm 1.1 \ {\rm MeV} 
\ee 
so we expect that a DK bound state with 
E$_{\rm B} \approx 40$~MeV would retain a significantly larger amplitude for 
$|{\rm B}\rangle = |{\rm D}^o {\rm K}^+\rangle$ 
than for
$|{\rm A}\rangle = |{\rm D}^+ {\rm K}^o\rangle$ 
in its state vector. This is equivalent to having
some admixture of
the symmetric 
$|{\rm I, I}_z\rangle = |1, 0\rangle$ DK state  
in addition to the dominant, antisymmetric
$|{\rm I, I}_z\rangle = |0, 0\rangle$ DK state.
The presence of an important $|{\rm I, I}_z\rangle = |1,0\rangle$ 
component in the dominantly I=0 DK molecule may account for 
the observed transition to D$_s^+\pi^o$.

\section{Prospects for additional molecules}

If the 2.32~GeV state seen by BaBar is indeed a
DK molecule, we might anticipate other heavy-quark molecular 
bound states in other channels that possess similar
attractive forces. In the Weinstein-Isgur binding mechanism
these are channels in which a $q\bar q$ state lies not far above
the two-meson continuum and has a strong decay coupling to 
S-wave meson pairs. 

There are many such possibilities. One that is rather similar to DK is
the channel D$^*$K, which has a threshold of 2.50~GeV. A broad
$c\bar s$ $1^+$ state which can provide attraction through
level repulsion is expected at 2.55~GeV \cite{God91}.  
The second BaBar signal, reported at a mass of 2.46~GeV \cite{Bab03}, 
is an obvious candidate for this molecular state; 
the mass difference of $2.46-2.32$~GeV can be understood as 
being essentially equal to
M(D$^*$)$-$M(D). (This assumes that the 
DK and D$^*$K binding energies
are comparable.) 
As an important test,
an S-wave D$^*$K molecule would have J$^{\rm P} = 1^+$.

The D$_s{\bar {\rm K}}$
system is analogous to DK in that mixing with $q\bar q$ 
intermediates is allowed, however in this case the important
mixing states are the {\it lighter} $c\bar n$ mesons, which are below
D$_s{\bar {\rm K}}$ threshold;
an effective 
D$_s{\bar {\rm K}}$ repulsion should result. 
Thus we would not expect molecular
states in this channel. Molecules with pions are also not expected, as they 
would have much smaller reduced masses that discourage the formation of
bound states.     

A state analogous to DK in the $c\bar b$  
system would be a BD
molecule, with a  B$_c^+ \pi$ decay mode that is  
isospin-conserving for I=1 
or isospin-violating for I=0.
As this is a heavy-light system, these states may more closely
resemble $Qqq$ baryons. 
Here too the masses are very different from the DK problem, and lead to 
a completely different
experimental signature, with a high energy pion. 
M(B) + M(D) $ \approx 7145$ MeV, whereas M(B$_c$) $\approx 6400 \pm 400$ MeV. 
So, a BD molecule just below threshold would 
simply rearrange the four quarks into
B$_c^+\pi$ and fall apart, either with or without isospin violation, giving a
neutral or charged pion having a well defined (but currently not well
determined) energy of 
$\approx 750\pm 400$~MeV,
with the precision improved by better measurements. Prospects 
for observing a relatively narrow BD molecule appear better for 
an I=0 state,
which involves an isospin-violating decay to B$_c^+\pi$.
This state might be observed as a resonance with a pion accompanying
the B$_c^+$, with an invariant mass too high to be confused with a
conventional $q \bar q$ state.

\section{Summary of experimental tests}

In summary: Challenges for experiment, which may help to determine the nature and dynamics
of this state, include: 
\begin{itemize}

\item
A better measure of the width to see if it may be much narrower than 10
MeV;

\item
A search for the mode 
D$_s^{*+} \pi$; the presence of D$_s^{+} \pi$ and 
absence of D$_s^{*+} \pi$ would uniquely select
J$^{\rm P} = 0^+$ (assuming strong or electromagnetic 
transitions);

\item
A search for the purely electromagnetic decay mode 
D$_s^+ \gamma$ (which is forbidden if the state is $0^+$) 
and the E1 transition to D$_s^{*+} \gamma$, to establish whether 
this partial width is markedly different from the 2~keV predicted for
a $c\bar s$ state;
 
\item
A search for charged partners appearing in 
D$_s^+ \pi^{\pm}$
that should exist if this is an isovector state;
 
\item 
Search for the $^3$P$_0$ D$_s(0^+)$ $c\bar s$ state 
with a mass of $\approx 2.5$~GeV; 
mass shifts relative to the D$_{sJ=1,2}$ partners may help quantify 
the dynamics leading to a DK bound state;
seek other possible narrow states below 2.36 GeV,
and determine their J$^{\rm P}$.

\item
Search in B decays for a possible D$\bar{\rm K}$ molecule,
to determine the dynamics of DK binding; one possible
signature could be D$^+{\rm K}^- \to {\rm K}^-{\rm K}^-\pi^+\pi^+$,
as in $\bar {\rm B}^o \to ({\rm D}^+ {\rm K}^-){\rm K}^o \to
 {\rm K}^-{\rm K}^-\pi^+\pi^+ {\rm K}^o$;
 
\item 
In $e^+e^-$ annihilation, 
measure the $Q^2$ dependence of the production cross section;
compare with the dependence observed for other charmed mesons
and with the counting rules for multiquark states; 
see if this dependence hardens at larger $Q^2$
due to a short range ``conventional" $c\bar{s}$ content; 
compare with the behavior of
$e^+e^- \to a_0(980)^+ a_0(980)^-$;

\item 
Precision data from CLEO-c in the 4.3-5 GeV region could
determine whether the threshold production process is 
$e^+e^- \to {\rm D}_{sJ}(2317)\bar{\rm D}_s^*(2112)$ 
in S-wave from $\sqrt{s} \geq 4.43$ GeV, 
or 
$e^+e^- \to 
{\rm D}_{sJ}(2317) {\bar {\rm D}_{sJ}(2317)}$
in P-wave,
from $\sqrt{s} \geq 4.64$ GeV; these can be compared with the 
threshold production of well-established charmed meson pairs;

\item 
If the ${\rm D}_{sJ}(2317)$ is indeed a DK 
molecule, search for further examples; 
there are many possibilities, including D$^*$K, and 
a BD molecule that might be observed in B$_c^+ \pi$.

\end{itemize}
 
\section{Acknowledgements}

We are indebted to A.Dzierba, S.Godfrey, R.L.Jaffe, 
D.Hitlin, V.Papadimitriou, 
J.Rosner, K.Seth, S.Spanier and J.Weinstein 
for useful discussions and communications. 

This research was supported in part by
the U.S. Department of Energy under contract
DE-AC05-00OR22725 at
Oak Ridge National Laboratory (ORNL), 
the Division of High Energy Physics, contract
W-31-109-ENG-38, 
by the
US-Israel Bi-National Science Foundation, and by 
the European Union under contract ``Euridice" HPRN-CT-2002-00311.  

\vfill\eject


\begin{thebibliography}{99}
\bibitem{Lin97} 
R.~Lindenbusch (Ph.D. thesis, Indiana University, 1997)
shows E852 data that is consistent with
$\pi\pi$ production of the nominally I=1 $a_0(980)$, which is  
identified in the $\pi\eta$ final state.
\end{thebibliography}
\end{document}